\documentclass{ws-ijmpcs}
\usepackage{url,overcite}
\DeclareMathOperator{\arccosh}{arccosh}
\begin{document}

\markboth{I. P. Lobo \& G. Palmisano}
{Geometric interpretation of Planck-scale-deformed co-products}

%
\catchline{}{}{}{}{}
%

\title{Geometric interpretation of Planck-scale-deformed co-products}

\author{Iarley P. Lobo}

\address{Dipartimento di Fisica, Sapienza Universit\`a di Roma, P.le Aldo Moro 5\\
Rome, 00185,
Italy\\
iarley.pereiralobo@icranet.org}

\author{Giovanni Palmisano}

\address{Dipartimento di Fisica, Sapienza Universit\`a di Roma, P.le Aldo Moro 5\\
Rome, 00185,
Italy\\
giovanni.palmisano1@gmail.com}

\maketitle

\begin{history}
\received{Day Month Year}
\revised{Day Month Year}
\published{Day Month Year}
\end{history}

\begin{abstract}
For theories formulated with a maximally symmetric momentum space we propose a general characterization for the description of interactions in terms of the isometry group of the momentum space. The well known cases of $\kappa$-Poincar\'e-inspired and (2+1)-dimensional gravity-inspired composition laws both satisfy our condition. Future applications might include the proposal of a class of models based on momenta spaces with anti-de Sitter geometry.
\end{abstract}

\ccode{PACS numbers: 03.30.+p, 02.40.-k, 04.60.-m}

\section{Introduction}	
\par
Doubly Special Relativity (DSR) is a ``bottom-up'' approach to the problem of describing the classical limit of quantum-gravitational regimes. The main feature of this formulation is the presence of an invariant energy scale, along with a deformed version of the kinematic equations of Special Relativity (SR), namely the dispersion relation and the conservation law for momenta, as well as a deformation of the Lorentz group of symmetries, in a way  such that the overall framework is compatible with a relativistic description.
\par
Such kind of description opens an important window for testing quantum-gravitational effects, which may be achievable by near future technology. Although part of the scientific community argues that such effects could be measurable only if we could directly access the Planck scale (for example, by the use of particle accelerators), recently, the possibility of having cosmological distances as natural amplifiers for probing Wheeler's ``space-time foam'' has inaugurated the area of Quantum-Gravity Phenomenology. For a review in the subject, see Ref.~\refcite{livingreviews}.
\par
The approaches named $\kappa$-Minkowski non-commutative space-time\cite{k-minkowski} and (2+1)-dimensional quantum-gravity \cite{2+1qg} are characterized by an algebraic structure, known as the co-product, which fixes the deformation of the conservation law in a way which is DSR-compatible\footnote{A formulation is DSR-compatible if its deformed Lorentz transformations preserves the form of the dispersion relation and preserves the deformed energy-momentum conservation law (possibly with back-reaction), i.e., $\Lambda(\xi, q\oplus p)=\Lambda(\xi,q)\oplus \Lambda(\xi \triangleleft q,p)$. This way, the conservation law that describes particle interactions is the same for all inertial observers. For details, see Ref.~\refcite{giulia}.} with the deformation of the symmetries and of the dispersion relation.
To this framework has been given recently a geometric characterization, known as ``Relative locality''\cite{relativelocality1, relativelocality2}, in which the deformation of the kinematics is interpreted in terms of a non-trivial geometric structure of momentum space. In particular, it has been realized that $\kappa$-Minkowski space-time and (2+1)-dimensional quantum-gravity can be related, respectively, to a de Sitter and an anti-de Sitter metric for the momentum space.\cite{kowalski}
\par
Such perception recovers Born's intuition,\cite{born} in which it is argued that to properly describe quantum effects of gravity, we should treat the momentum space and space-time on ``equal-footing'', just like in the case of quantum mechanics. Therefore, from these considerations, in order to incorporate gravitational effects in quantum mechanics, a suggestive way would be to use a description where the momentum space has some curvature.
\par
In this paper we identify and analyze to some extent a peculiar geometric characterization of the relation between the dispersion relations and the conservation laws of both $\kappa$-Poincar\'e and (2+1)-dimensional quantum-gravity momentum spaces, having a role which resemble the one played by the co-product  in the group theoretical setting in which both spaces where developed.
\par
This paper is organized as follows. In section \ref{section2} we review the most recent formalization of momentum space as a curved manifold and some of its possible affine connections. In section \ref{section3} we intrinsically define our prescription for a relativistic composition law as the translations in maximally symmetric spaces. In section \ref{section4} we explicitly investigate the non-trivial cases of (anti-) de Sitter metrics and find some stablished momentum spaces presented in the quantum-gravity literature, implying that our prescription, which is grounded on a pure geometric intuition, is the geometric interpretation of the co-product and may serve for further investigations in the Relative Locality formalism. Finally, in section \ref{conclusion} we draw some perspectives for future applications of our construction. The results here reported are part of a larger project which should be completed soon. \cite{mainarticle}

\section{Relative Locality}\label{section2}
\par
In the context of Special Relativity we can use space-time coordinates to label events because from its axioms the speed of light is source independent and a universal constant. Besides this condition, we need also to require that the detection and emission of light signals should be frequency independent. This way, all observers are able to construct the same space-time by means of clock synchronization and radar experiments from the exchange of light signals. In the ``Relative Locality'' approach it is exhibited that such idealization is due to the fact that momentum space is considered a flat manifold. Therefore, if one assumes that momentum space has the non-trivial structure of a general differential manifold, it is possible to generalize Einstein's construction of a space-time and derive that locality is not an absolute (observer independent) concept, but a relative one. \cite{relativelocality1, relativelocality2}
\par
The way in which the Relative Locality framework interprets the deformed kinematics in terms of a non-trivial geometric structure for the momentum space consists in associating a metric to the dispersion relation and a connection to a conservation law. In fact, the mass $m_p$ of a particle with momentum $p$ is defined as the geodesic distance from a point chosen as the origin $\underline{0}$ to a point with coordinate $p$
\begin{equation}
m_p^2=d^2(p,\underline{0})=\int_{\underline{0}}^{p} dt\sqrt{g^{\mu\nu}\dot{\gamma}_{\mu}\dot{\gamma}_{\nu}},
\end{equation}
where $\gamma$ is the geodesic connecting points $\underline{0}$ and $p$ that depends only on the Christoffel symbols of the metric. Using the mikowskian metric, Special Relativity's dispersion relation is a special case of this one, where the geodesic is a straight line from the origin.
\par
The composition law $q\oplus p$ that governs the interaction of particles with momentum $q$ and $p$ is related to the connection\footnote{The connection arising from formula (\ref{connection}) will be, in general, different from the Levi-Civita connection associated to the given metric, it may have a non-vanishing torsion or non-metricity.} at a point $k$ by
\begin{equation}
\Gamma^{\mu\nu}_{\alpha}(k)=-\frac{\partial}{\partial q_{\mu}}\frac{\partial}{\partial p_{\nu}}(q\oplus_{k} p)_{\alpha}|_{q=p=k}\label{connection},
\end{equation}
where $\oplus_{k}$ is the ``translated'' composition law
\begin{equation}
q\oplus_{k}p=\oplus k(\ominus k\oplus q)\oplus (\ominus k\oplus p)),
\end{equation}
and $\ominus$ is the, so called antipode operation of $\oplus$. It is such that
\begin{equation}
(\ominus p)\oplus p=0=p\oplus (\ominus p),
\end{equation}
and $\oplus$ is $\oplus_k$ for $k=\underline{0}$.
\par
In Ref.~\refcite{palmisano} it is presented a second alternative for the relation between the composition law and the connection given by the prescription of associating any couple of momenta $q$ and $k$ to the autoparallel\footnote{Those are curves $\gamma$ such that $\frac{d^{2}\gamma_{\lambda}}{dt^2}+\Gamma_{\lambda}^{\mu\nu}\left(\gamma\right)\frac{d\gamma_{\mu}}{dt}\frac{d\gamma_{\nu}}{dt}=0$.} geodesics $\gamma^{(q)}$ and $\gamma^{(k)}$ which connect them to the
origin of momentum space. A parametric surface $\gamma\left(s,t\right):[0,1]\times[0,1]\rightarrow M$
function of the parameters $s$ and $t$ is then defined on the momentum
space $M$ by:
\begin{subequations}
\begin{align}
\frac{\partial}{\partial t}\frac{\partial}{\partial s}\gamma_{\lambda}+\Gamma_{\lambda}^{\mu\nu}\left(\gamma\right)\frac{\partial\gamma_{\mu}}{\partial t}\frac{\partial\gamma_{\nu}}{\partial s}=0,\\
\gamma_{\lambda}\left(s,0\right)=\gamma_{\lambda}^{(k)}\left(s\right),\\
\gamma_{\lambda}\left(0,t\right)=\gamma_{\lambda}^{(q)}\left(t\right).
\end{align}
\end{subequations}
\par
The composition of momenta, in the end, is interpreted as the point
given by:
\begin{equation}
q\oplus k=\gamma\left(1,1\right).
\end{equation}
\par
Although these two interpretations of the composition law are in general
different, in the following we will consider cases in which they are equivalent.\footnote{This is is always the case, as showed in Ref.~\refcite{palmisano}, when the composition
law is associative.}

\section{Definition of the $\kappa$-Composition Law}\label{section3}
\par
Let's define what we call a $\boldsymbol\kappa$\textit{-composition law}:\footnote{The same constrains, although in a more specific framework, were also considered in Ref.~\refcite{freidel2}.}
\begin{subequations}\label{kappacomp}
\begin{align}
d\left(q\oplus p,q\oplus k\right)=d\left(p,k\right),\\
q\oplus p|_{q=0}=p,\\
q\oplus p|_{p=0}=q.
\end{align}
\end{subequations}
\par
As $d(p,k)$ is the riemannian geodesic distance from $p$ to $k$, this definition means that the map $p\mapsto q\oplus p$, is generated by Killing vectors in the manifold, with parameters given by the components of momentum $q$. To specialize the geometry of momentum space, in order to describe a DSR theory we should consider maximally symmetric spaces. Working just on Minkowski, de Sitter and anti-de Sitter metrics, it is well known that the Killing equations admit the maximal number of solutions, in four dimensions we have ten Killing vectors. Transformations generated by these vectors that preserve the origin are interpreted as deformed Lorentz transformations, including boosts and rotations. Those that are inhomogeneous, i.e., satisfy requirements (\ref{kappacomp}) are responsible for a deformed energy-momentum conservation law and describe a deformed picture for interactions. For a minkowskian momentum space the analysis essentially leads to Special Relativity; novel features come supposing (anti-) de Sitter spaces.\footnote{The constant curvature of these spaces is given by an energy scale $\kappa$. This is the origin of our composition law's name.} As will be described in the next section, de Sitter case leads simply to another well known formalism, the one of $\kappa$-Poincar\'e momentum space. For anti-de Sitter we recover the momentum space of (2+1)-dimensional quantum-gravity. 
\par
Besides the mass invariance under symmetry maps, the DSR-compatibility condition means that the result of a Lorentz transformation on composed momenta is the composition of individual Lorentz-transformed momenta, i.e., roughly speaking, there should be a well-defined commutation rule between the composition operation and Lorentz maps. As the $\kappa$-composition law is derived from the isometry group of the manifold, it naturally has such defined commutation relation with the homogeneous sector of the isometry group that defines the deformed Lorentz transformations. Such property is present in well-known formalisms that satisfy our prescription for a composition law, like $\kappa$-Poincar\'e and (2+1)-dimensional quantum-gravity inspired momentum spaces, as well as the one from Ref.~\refcite{freidel2} related to Snyder momentum space.
\par
The $\kappa$-composition law, defined by (\ref{kappacomp}), is a kind of compatibility condition between the metric and the connection. Knowing the dispersion relation we can guess the metric and with conditions (\ref{connection}), (\ref{kappacomp}) we can find the connection from the isometries of the manifold.
\par
It is important to notice that the $\kappa$-composition law is defined up to Lorentz transformations. A more detailed analysis of such ambiguity will be discussed in the Ref.~\refcite{mainarticle}.

\section{Applications}\label{section4}

\subsection{De Sitter momentum space}
A useful method one can choose to analyse de Sitter space consists in embedding it in a flat, higher-dimensional manifold.\cite{moschella} We use Greek letters to label coordinates from $0,...,3$. One can describe a $4$ dimensional de Sitter manifold as a hypersurface embedded in a $5$ dimensional Minkowski space with signature $(+----)$, with an extra space-like coordinate. For a minkowskian metric 
\begin{equation}
ds^2=dP_0^2-dP_1^2-dP_2^2-dP_3^2-dP_4^2,
\end{equation}
de Sitter space is defined as the hyperboloid satisfying 
\begin{equation}
P_0^2-P_1^2-P_2^2-P_3^2-P_4^2\doteq |P|^2-P_4^2=-\kappa^2.
\end{equation}
\subsubsection{Comoving coordinates}
\par
A possible choice of coordinate system for de Sitter space is
\begin{subequations}\label{comoving}
\begin{align}
P_0=\kappa \sinh(p_0/\kappa)+e^{p_0/\kappa}(p_1^2+p_2^2+p_3^2)/2\kappa,\\
P_i=e^{p_0/\kappa} p_i,\\
P_4=\kappa\cosh(p_0/\kappa)-e^{p_0/\kappa}(p_1^2+p_2^2+p_3^2)/2\kappa.
\end{align}
\end{subequations}
\par
These are called comoving coordinates and the metric assumes the following form:
\begin{equation}
ds^2=dp_0^2-e^{2p_0/\kappa}d\Sigma^2,
\end{equation}
where $d\Sigma^2$ is the flat spatial line element. This is the coordinate system usually used to describe $\kappa$-Poincar\'e momentum space\cite{giulia}. It covers only half of de Sitter space, the portion $P_0+P_4>0$.
\par
The dispersion relation of a particle with mass $m_p$ and momenta $(p_0,p_i)$ is defined as the geodesic distance $d$ from the origin $\underline{0}=(0,0)$ to $(p_0,p_i)$. In Ref.~\refcite{giulia} it is shown that
\begin{equation}
m_p=d=\kappa\arccosh(P_4/\kappa)=\arccosh\left[\cosh(p_0/\kappa)-e^{p_0/\kappa}(p_1^2+p_2^2+p_3^2)/2\kappa^2\right].\label{on-shellds}
\end{equation}
\par
Using our prescription of a $\kappa$-composition law, solving the Killing equations and integrating the infinitesimal parameters to get a finite transformation, we find a possible solution:
\begin{subequations}\label{compsitter}
\begin{align}
(q\oplus p)_0=q_0+p_0,\\
(q\oplus p)_i=q_i+p_ie^{-q_0/\kappa}.
\end{align}
\end{subequations}
\par
These dispersion relation and composition law are exactly the ones of  the $\kappa$-Poincar\'e momentum space,\cite{giulia} that are usually derived from group theoretical definitions.\cite{kowalski} So, we were able to reconstruct the $\kappa$-Poincar\'e momentum space using purely geometric tools and intuition.
\par
In these coordinates the connection is not the Levi-Civita one because its curvature tensor is null. This was already expected, because as the $\kappa$-composition law is, essentially, an isometry, this implies that it is associative, and from (\ref{connection}) we can see that the Riemann tensor is a measure of its associativity.
\par
In the five-dimensional-embedding setting, we can write such isometry as a linear transformation
\begin{equation}
\textbf{Q}\oplus \textbf{P}=\textbf{T}(Q)_{dS}\textbf{P},
\end{equation}
such that
\begin{eqnarray}
\textbf{T}(Q)_{dS}=
\left(
  \begin{array}{ccccc}
    \frac{Q_0}{\kappa}+\frac{\kappa}{Q_0+Q_4} & \frac{Q_1}{Q_0+Q_4} & \frac{Q_2}{Q_0+Q_4} & \frac{Q_3}{Q_0+Q_4} & \frac{Q_0}{\kappa} \\
    Q_1/\kappa & 1 & 0 & 0 & Q_1/\kappa \\
    Q_2/\kappa & 0 & 1 & 0 & Q_2/\kappa \\
    Q_3/\kappa & 0 & 0 & 1 & Q_3/\kappa \\
    \frac{Q_4}{\kappa}-\frac{\kappa}{Q_0+Q_4} & -\frac{Q_1}{Q_0+Q_4} & -\frac{Q_2}{Q_0+Q_4} & -\frac{Q_3}{Q_0+Q_4} & \frac{Q_4}{\kappa} \\
  \end{array}
\right).
\end{eqnarray}
\par
This matrix is an element of $SO(4,1)$, it preserves the five-dimensional minkowskian metric $\boldsymbol\eta=diag(+----)$, i.e., $\textbf{T}\boldsymbol\eta\textbf{T}^T=\boldsymbol\eta$.
\par
This portion of de Sitter space is known to be a group manifold. In the group theoretical analysis of non-commutative $\kappa$-Minkowski space-time, it is claimed that its momentum space is isomorphic to de Sitter space, this way, one can use the notion of {\it co-product} \cite{kowalski} to define a composition rule for two particles involved in an interaction. As the total momentum of the resulting interaction is still an element of the group, composition laws constructed from co-products are equivalent to our geometric $\kappa$-composition law. We were able to derive these algebraic results using a pure geometric intuition.

\subsection{Anti-de Sitter momentum space}
\par
The anti-de Sitter metric can be found using a similar procedure of embedding a hypersurface in a flat space.\cite{moschella} Let's do the example of embedding the three dimensional anti-de Sitter space in a four-dimensional one with signature $(+--+)$. For the minkowskian metric
\begin{equation}
ds^2=dP_0^2-dP_1^2-dP_2^2+dP_4^2,
\end{equation}
anti-de Sitter space is defined as the hyperboloid satisfying 
\begin{equation}
P_0^2-P_1^2-P_2^2+P_4^2=\kappa^2.
\end{equation}
\subsubsection{Cartesian coordinates}
A possible choice of coordinates for anti-de Sitter space is
\begin{subequations}\label{cartesian2}
\begin{align}
P_{\mu}=p_{\mu}\ \ , \ \ \mu=0,...,2,\\
P_4=\sqrt{\kappa^2-|p|^2},
\end{align}
\end{subequations}
the metric assumes the following form
\begin{equation}
ds^2=dp_{\mu}dp_{\nu}\left(\eta^{\mu\nu}+\frac{p^{\mu}p^{\nu}}{\kappa^2-|p|^2}\right),
\end{equation}
where $\eta^{\mu\nu}=diag(+--)$ is the three-dimensional minkowskian metric. This coordinate system also covers only half of anti-de Sitter space, the one defined by $P_4>0$.
\par
As before, the dispersion relation of a particle with mass $m_p$ and momenta $p$ is defined as the geodesic distance from the origin to $p$. This way, we are able to recover the result from Ref.~\refcite{spinning}
\begin{equation}
m_p=d=\kappa\arccos(P_4/\kappa)=\kappa\arcsin\left(|p|/\kappa\right).
\end{equation}
\par
Furthermore, it is straightforward to verify that the following equation, derived using group theoretical considerations,\cite{hopf} is a $\kappa$-composition law
\begin{equation}
(q\oplus p)_{\mu}=\kappa^{-1}(q_4p_{\mu}+p_4q_{\mu}+\epsilon_{\mu}^{\ \ \alpha\beta}q_{\alpha}p_{\beta}),
\end{equation}
where $\epsilon_{\mu}^{\ \ \alpha\beta}$ is the minkowskian Levi-Civita symbol and $\epsilon_{012}=1$.
\par
In the four-dimensional-embedding setting, the composition law is performed by the matrix
\par
\begin{eqnarray}
\textbf{T}(Q)_{ads}=
\kappa^{-1}\left(
  \begin{array}{ccccc}
    Q_4 & -Q_2 & Q_1 & Q_0 \\
    -Q_2 & Q_4 & Q_0 & Q_1 \\
    Q_1 & -Q_0 & Q_4 & Q_2 \\
    -Q_0 & Q_1 & Q_2 & Q_4 \\
  \end{array}
\right),
\end{eqnarray}
it is straightforward to verify that $\textbf{T}\boldsymbol\eta\textbf{T}^T=\boldsymbol\eta$ for $\boldsymbol\eta=diag(+--+)$, i.e., it is an element of $SO(2,2)$.
\par
As in the previous case, this is the momentum in which particles are defined in (2+1)-dimensional quantum-gravity. The ``spinning'' composition law that we wrote above is exactly the one extracted using the definition of co-product in the portion of anti-de Sitter space we considered, which is a group manifold.\cite{2+1qg, hopf}

\section{Conclusion}\label{conclusion}
In this paper, we proposed an intrinsic, DSR-compatible rule that prescribes how momenta of particles could be composed if their momentum space has a non-trivial structure of a differential manifold endowed with a maximally symmetric metric and a non-riemannian affine connection. Using this prescription, we recovered the $\kappa$-Poincar\'e-inspired and (2+1)-dimensional quantum-gravity-inspired momentum spaces, which are known in the quantum-gravity community.  The former is described by a de Sitter metric, while the latter is described by an anti-de Sitter metric in a three-dimensional manifold.
\par
In this curved momentum space paradigm, de Sitter geometry has been extensively studied in the literature for any number of dimensions (for a review, see Ref.~\refcite{kowalski}). But a persistent, open problem is the description of a phenomenology based on the four-dimensional anti-de Sitter geometry. Such analysis is important because it may predict effects that are not achievable by a de Sitter geometry. 
\par
A first attempt in this direction was done in Ref.~\refcite{arzano} for the one-particle case. However, it is still missing an analysis for the multi-particle system due to the lack of a rule that governs how momenta compose themselves. The momentum spaces we here discussed, inspired by the theoretical background of $\kappa$-Minkowski space-time and (2+1)-dimensional quantum-gravity are, in a sense, unified around our geometric concept of a $\kappa$-composition law, and our proposal may, then, serve as a guiding principle towards the derivation of a coherent four-dimensional anti-de Sitter momentum space. This paper is part of large project\cite{mainarticle} that aims to furnish such coherent picture for the community of quantum-gravity phenomenology.

\section*{Acknowledgments}

IPL acknowledges the support by the CAPES-ICRANet program financed by CAPES - Brazilian Federal Agency for Support and Evaluation of Graduate Education within the Ministry of Education of Brazil through the grant BEX 14632/13-6.


\end{document}